\documentclass[11pt]{article}
\begin{document}
\begin{center}
\today \\
\begin{Large} {\bf   Thermal Effects of Rotation in  Random Classical
Zero-Point Radiation. }
\end{Large}
\vspace{10mm}\\
\begin{it}
{Yefim S. Levin }\\
Department of Electrical and Computer Engineering, Boston University,
Boston MA, 02215   \\
\end{it}
\end{center}
\begin{abstract}
\indent The rotating reference system $\{\mu_\tau\}$, along with
the two-point correlation  functions (CFs) and energy density, is
defined and used as the basis for investigating  thermal effects
observed by  a detector rotating through random classical
zero-point radiation. The reference system consists of Frenet
-Serret orthogonal tetrads
 $\mu_\tau$. At each proper time $\tau$ the rotating detector is
 at rest  and has a constant acceleration vector at the
 $\mu_\tau$.
 \\
\indent The two-point CFs and the energy density at the rotating
reference system should be periodic with the period $T=\frac{2
\pi}{\Omega}$, where $\Omega$ is an angular detector velocity,
because CF and energy density measurements is one of the tools the
detector can use to justify the periodicity of its motion. The CFs
 have been calculated  for both \emph{electromagnetic} and
 \emph{massless scalar} fields in two cases, with and without taking
  this periodicity into
consideration. It turned out that only periodic CFs have some
thermal features and  particularly the Planck's factor with the
temperature $T_{rot}=\frac{\hbar \Omega}{2 \pi k_B}$ ($k_B$ is the
Boltzman constant).  \\
\indent Regarding to the energy density of both electromagnetic
and massless scalar field it is shown that the detector   rotating
in the zero-point radiation observes not only this original
zero-point radiation but , above that, also the radiation
 which would
have been observed by an inertial detector in the thermal bath
with the Plank's spectrum at the temperature $T_{rot}$. This
effect is masked by  factor $\frac{2}{3}(4\gamma^2-1)$ for the
electromagnetic field and $\frac{2}{9}(4\gamma^2-1)$ for the
massless scalar field, where $\gamma=(1-(\frac{\Omega
r}{c})^2)^{-1/2}$. Appearance of these masking factors is
connected with the fact that rotation is defined by two
parameters, angular velocity and the radius of rotation, in
contrast with a uniformly accelerated linear motion which is
defined by only one parameter, acceleration a. \\
\indent Our calculations involve classical point of view only and
to the best of our knowledge the results have not been reported in
quantum theory yet.
\end{abstract}
\section{Introduction}.
\indent This article is focused on investigations of the thermal
effects connected with rotation of a classical detector moving
through a random classical zero-point field. The case of a linear
uniformly accelerated detector
 has been investigated and used in many
works \cite{boyer1980, boyer1984, heish1994, cole1986}. The
rotation , to the best of our knowledge, has not been considered
in random \emph{classical} radiation.  It was studied in
\emph{quantum} case and only for the massless scalar field,
particularly in connection with rotating vacuum puzzle
\cite{davis1996, lorency2000, suga1999} and different
coordinate mappings.\\
\indent  Our approach in CF calculation is very close but not
identical to the method developed in \cite{boyer1980, boyer1984}
for uniformly accelerated detectors and also comes from the ideas
developed in \cite{irvine}
and\cite{synge}. \\
\indent The observations made by the rotating detector are
described in the non-inertial reference system consisting of
instantaneous inertial reference frames and mathematically defined
as tetrads at each moment of the detector proper time. Along with
such reference system the two-point correlation functions of the
electromagnetic and scalar massless field and one-point energy
density of these fields are defined and analyzed for
zero-point radiation.\\
\indent The article is organized as follows. In  section
\ref{sec:TetradsMeasuments} we  explicitly show the relationship
between the components of the electromagnetic field measured at a
tetrad and in the laboratory coordinate system. In section
\ref{sec-Definition} the Frennet-Seret orthogonal tetrads
connected with a rotating point-like detector are defined and its
merits for our problems compared with the Fermi-Walker tetrads are
discussed. The CFs of the electromagnetic field at the rotating
detector moving through the classical zero-point electromagnetic
filed are considered in  section
\ref{sec-CorFunForEelctromagneticFieldAtRot} and their expressions
are presented in terms of elementary functions. In  section
\ref{sec-classicalSpectrum} these CFs and the energy density
measured by the rotating detector are analyzed taking into
consideration periodicity of these functions and then using the
Abel-Plana summation function. Due to the periodicity of the
motion , the observer rotating through a zero-point
electromagnetic radiation  sees the energy density, which would
have have been observed by  an inertial observer moving in a
thermal bath at the temperature $T_{rot}=\frac{\hbar \Omega}{2 \pi
k}$, and multiplied by the factor $\frac{2}{3}(4 \gamma^2 -1)$.
For the classical massless scalar zero-point field the CF,  energy
density, and their thermal properties connected with the
periodicity are investigated in the section \ref{sec-massless}.
Calculation details are presented in Appendix.
\section{ Tetrads and Measurements by a Rotating Detector  }.
\label{sec:TetradsMeasuments}
 \indent We will start from the case of random classical
\emph{electromagnetic} field. A local inertial system of a
rotating point-like detector (observer )  is associated with an
orthogonal tetrad $\mu$ of four vectors $\mu^i_{\; (a)}, \;\;
a=1,2,3,4, $ defined at its location. The tensor of
electromagnetic field, $F_{i k}$, defined at this point, may be
resolved  along these  frame vectors   according to the following
formula \cite{irvine}
\begin{eqnarray}
F_{ik}=\mu^{(a)}_i \; \mu^{(b)}_k \; F_{(ab)}(\mu).
\end{eqnarray}
The components
\begin{eqnarray}
\label{eq:FrameTensor} F_{(ab)}(\mu)=\mu^{i}_{(a)} \; \mu^k_{(b)}
\; F_{i k}.
\end{eqnarray}
 are scalars under coordinate transformations. But
they clearly transform upon change of orthogonal tetrads $\mu$
according to the local (that is defined at the same point) Lorentz
transformations. The electric and magnetic fields, $E_{(a)}$ and
$H_{(b)}$, measured by a rotating detector are defined in terms of
the components $F_{(ab)}$:
\begin{eqnarray}
\label{eq:FrameField} (E_{(a)}(\mu))=(F_{(41)}(\mu),
F_{(42)}(\mu),F_{(43)}(\mu)), \;\; (H_{(a)}(\mu))=F_{(23)}(\mu),
F_{(31)}(\mu),F_{(12)}(\mu)).
\end{eqnarray}
They do not depend on the coordinate system and are connected with
a tetrad. \\
\indent If  the tetrad vectors $\mu^i_{\; (a)}, \;\; a=1,2,3,4, $
of a reference frame $\mu$  and the electromagnetic field state
$F_{ik}$ are specified in the laboratory coordinate system then
according to (\ref{eq:FrameTensor}) the electric and magnetic
fields at a rotating detector can be expressed in terms of the
electric and magnetic fields in the laboratory coordinate
system. It will be done in the next sections to calculate CFs.\\
\section{The Definition of the Local Reference Frame (Tetrad)
for a Rotating Detector}. \label{sec-Definition} \indent There are
different ways to define orthogonal tetrads  \cite{synge}
,\cite{ivanitskaya}. Tetrads with the time component proportional
to 4-vector velocity of an observer define reference frames in
which the observer is at rest. There are two types of such
tetrads. The so called not rotating tetrads are described by the
Fermi-Walker equations and are used to explain the Thomas
procession \cite{moller}. Frenet-Serret orthogonal tetrads give an
example of so called rotating tetrads and may also be used for
observations \cite{misner73}. \\
\indent In this publication the reference frames $\mu_\tau$ are
defined as Frenet-Serret orthogonal tetrads on the world line of
the rotating detector . For the rotating detector, with 4-vector
velocity
\begin{eqnarray}
U^i=c\;(-\beta \gamma \sin \alpha,\beta \gamma \cos \alpha, 0,
\gamma), \;\; \alpha=\Omega \gamma \tau, \;\;
\gamma=(1-\frac{(\Omega a)^2}{c^2})^{1/2},
\end{eqnarray}
 the four vectors of a Frenet-Serret orthogonal
tetrad have been found to be (Appendix \ref{sec:FrenetSerret})
\begin{eqnarray}
\label{eq:FStetrad}
\mu^i_{(1)}=(\cos \alpha, \sin \alpha, 0,0), \nonumber \\
\mu^i_{(2)}=(-\gamma \sin \alpha, \gamma \cos \alpha,0, \gamma).
 \nonumber \\
 \mu^i_{(3)}=(0,0,1,0), \nonumber \\
 \mu^i_{(4)}=\frac{U^i}{c}.
\end{eqnarray}
In the local reference frames, defined by these tetrads, the
detector is at rest at any proper time  $\tau$
because
\begin{eqnarray}
U_{(a)}=\mu^i_{(a)}U_i= \mu^i_{(a)} U^k g_{ik}= (0,0,0,-c).
\end{eqnarray}
and 4-vector acceleration of the detector in it
\begin{eqnarray}
\dot{U}_{(a)}= \mu^i_{(a)}\dot{U}_i=\mu^i_{(a)}\dot{U}^k g_{ik}
=(-a \Omega^2 \gamma^2,0,0,0), \;\;\; g_{ik}=diag(1,1,1,-1)
\end{eqnarray}
is constant in both magnitude and direction
\begin{eqnarray}
\dot{U}_{(a)}=(-a \Omega^2 \gamma^2,0,0,0)=(-a \Omega^2
\gamma^2,0,0,0).
\end{eqnarray}
  \indent
  Unlike Frenet-Serret tetrads  Fermi-Walker tetrads do not have this
   feature. In  the reference
frame associated with a Fermi-Walker tetrad  the 4-vector
acceleration is not constant in neither direction nor magnitude
\begin{eqnarray}
\label{eq:FermiWalkerAcceleration} \dot{U}_{(a)}=
e_{(a)l}\dot{U}_l= (-a \Omega^2\gamma^2 \cos \alpha \gamma,-a
\Omega^2 \gamma^2 \sin \alpha \gamma,0,0).
\end{eqnarray}
( Fermi-Walker's tetrad vectors $e_{(a)}$ are given in  Appendix
\ref{sec:FrenetSerret}).
The acceleration depends on $\tau$ and at each $e_{\tau}$ is different. \\
\indent  This is why we use  Frenet-Serret  tetrads for the case
of the rotation, not Fermi-Walker ones. In this formalism rotation
is becoming similar to a linear uniformly accelerated motion, with
a constant acceleration vector in an instantaneous rest reference
frame.
\section{Correlation Functions for Electromagnetic Field at a Rotating
Detector }\label{sec-CorFunForEelctromagneticFieldAtRot}
\subsection{Electric and Magnetic Fields in a
Frenet-Serret tetrad}
 \indent Following  formulas (\ref{eq:FrameTensor},
 \ref{eq:FrameField}, \ref{eq:FStetrad}),  the electric
 $ E_{(k)}(\mu | \tau)$ and magnetic  $H_{(k)}(\mu | \tau)$
 fields  in
the Frenet-Serret reference frame  $\mu_{\tau}$ at the proper time
$\tau$ of the rotating detector
 can be given in
terms of electric  and magnetic fields in the laboratory
coordinate system:
\begin{eqnarray}
\label{eq:FieldAtTetrad} E_{(1)}(\mu | \tau)=F_{(41)}(\mu | \tau)=
E_1 \gamma \cos \alpha + E_2\gamma \sin \alpha
-H_3 \beta \gamma, \nonumber \\
E_{(2)}(\mu | \tau)=F_{(42)}=E_1(-\sin \alpha) +
E_2\cos \alpha, \nonumber\\
E_{(3)}(\mu | \tau)=F_{(43)}= E_3 \;\;\gamma + H_1\beta \gamma
\cos\alpha +
H_2 \beta \gamma \sin \alpha,  \nonumber \\
H_{(1)}(\mu | \tau)=F_{(23)}= H_1\gamma \cos \alpha + H_2\gamma
\sin \alpha
+ E_3\beta \gamma,  \nonumber \\
H_{(2)}(\mu | \tau)=F_{(31)}= H_1(-\sin \alpha) +
H_2 \cos \alpha,\nonumber  \\
H_{(3)}(\mu | \tau)=F_{(12)}= H_3\gamma + E_1(\beta \gamma \cos
\alpha) + E_2(-\beta \gamma \sin \alpha),
 \end{eqnarray}
 where $\gamma=1/(1 -\beta^2)^{1/2}, \;\; \beta=v/c, \; \;
 v=\Omega a, \; \; \alpha =\Omega \gamma \tau =\Omega t$,
 $\Omega$ is an angular velocity of the rotating detector, and a
 is the radius of the rotation circle. \\
\indent In these equations $E_k$ and $H_k$ are the electric and
magnet field components of the random zero-point radiation in the
laboratory coordinate system at the location of the rotating
detector at its proper time $\tau$ \cite{boyer1980}:
\begin{eqnarray}
\label{eq:FieldOnCircle} \vec{E}(\tau)= \sum^2_{\lambda=1} \int
d^3k
\hat{\epsilon}(\vec{k},\lambda)h_0(\omega)\cos[\vec{k}\vec{r}(\tau)-
\omega
\gamma \tau -\Theta(\vec{k},\lambda)], \nonumber \\
\vec{H}(\tau)=\sum^2_{\lambda=1} \int d^3k
[\hat{k},\hat{\epsilon}^(\vec{k},\lambda)]h_0(\omega)\cos[\vec{k}
\vec{r}(\tau)-\omega \gamma \tau-\Theta(\vec{k},\lambda)],
\label{eq:ff}
\end{eqnarray}
where the $\theta(\vec{k},\lambda)$ are random phases distributed
uniformly on the interval $(0,2\pi)$ and independently for each
wave vector $\vec{k}$ and polarization $\lambda$ of of a plane
wave, and
\begin{eqnarray}
 \pi^2 h_0^2(\omega)=(1/2)\hbar \omega, \;\;\; \vec{r}(\tau)=
 (a \cos \Omega \gamma \tau, a \sin \Omega \gamma \tau, 0).
 \end{eqnarray}
 \indent Using formulas (\ref{eq:FieldAtTetrad}) we can calculate
 all two-field correlation functions
 $\langle E_{(a)}(\mu_1|\tau_1)E_{(b)}(\mu_2|\tau_2)\rangle$,
 $\langle E_{(a)}(\mu_1|\tau_1)H_{(b)}(\mu_2|\tau_2)\rangle$, and
 $\langle H_{(a)}(\mu_1|\tau_1)H_{(b)}(\mu_2|\tau_2)\rangle$, $a,b=1,2,3$
 of the
 electromagnetic field at the rotating detector. Here $\langle \rangle$
  means averaging over random phases $\theta(\hat{k},
 \lambda)$ \cite{boyer1980} We will start
 with $\langle E_{(1)}(\mu_1|\tau_1)E_{(1)}(\mu_2| \tau_2)\rangle$.
\subsection{ The Correlation Function
$\langle E_{(1)}(\mu_1|\tau_1)E_{(1)}(\mu_2| \tau_2)\rangle$:
General Expression  } \label{sec-general}
 \indent As an example let us  first treat
with
\begin{eqnarray}
I_{(11)}=\langle E_{(1)}(\mu_1|\tau_1)E_{(1)}(\mu_2|
\tau_2)\rangle.    \label{eq: CF}
\end{eqnarray}
In this expression $\mu_1$ and $\mu_2$ are two reference frames
(tetrads) on the circle of the rotating detector at the proper
times $\tau_1$ and $\tau_2$ respectively. Then
 from
 (\ref{eq:FieldAtTetrad}) follows
\begin{eqnarray}
I_{(11)}= <E_1(\tau_1)E_1(\tau_2)> \gamma^2 \cos \alpha_1 \cos
\alpha_2 + <E_1(\tau_1)E_2(\tau_2)> \gamma^2 \cos \alpha_1 \sin
\alpha_2 +  \nonumber \\
<E_2(\tau_1) E_1(\tau_2)> \gamma^2 \sin \alpha_1 \cos \alpha_2 +
<E_1(\tau_1)H_3(\tau_2)> (-1)\gamma^2 \cos \alpha_1 +
\nonumber \\
<H_3(\tau_1)E_1(\tau_2)>(-1)\beta \gamma^2 \cos \alpha_2 +
<E_2(\tau_1)E_2(\tau_2)> \gamma^2 \sin \alpha_1 \sin \alpha_2 +
\nonumber \\
<E_2(\tau_1)H_3(\tau_2)>(-1) \beta \gamma^2 \sin \alpha_1 +
<H_3(\tau_1)E_2(\tau_2)>(-1)\beta \gamma^2 \sin \alpha_2 +
\nonumber \\
<H_3(\tau_1)H_3(\tau_2)>
\end{eqnarray}
The $<>$ expressions can be calculated using
(\ref{eq:FieldOnCircle}),  the relationships
\begin{eqnarray} \label{eq:thetas}
<\cos \theta(\vec{k}, \lambda)\cos \theta(\vec{k}^\prime,
\lambda^\prime)>=<\sin \theta(\vec{k}, \lambda)\sin
\theta(\vec{k}^\prime, \lambda^\prime)>= \frac{1}{2}
\delta_{\lambda \;\lambda^\prime}
\delta^3(\vec{k}-\vec{k}^\prime), \nonumber \\
 <\cos
\theta(\vec{k}, <\cos \theta(\vec{k}, \lambda)\sin
\theta(\vec{k}^\prime, \lambda^\prime)>=0, \nonumber \\
\sum^2_{\lambda=1}\epsilon_i(\vec{k},
\lambda)\epsilon_i(\vec{k}^\prime,
\lambda)^\prime=\delta_{ij}-k_i\;k_j/k^2=\delta_{ij}-\hat{k}_i
\hat{k}_j,
\end{eqnarray}
from \cite{boyer1980} and variable change in the integrands
\begin{eqnarray}
\label{eq:VariableChange} \hat{k}_x \cos \alpha + \hat{k}_y \sin
\alpha =\hat{k^\prime}_x, \;\; -\hat{k}_x \sin \alpha + \hat{k}_y
\cos \alpha =\hat{k^\prime}_y, \;\;\alpha=\frac{\alpha_1 +
\alpha_2}{2}=\frac{\Omega \gamma (\tau_2+\tau_1)}{2}, \\
\hat{k}_i=k_i/k, \;\; i=x, y, z.
\end{eqnarray}
In the final integrand  the prime symbol of the dummy variable
$k^{\prime}$ is omitted for simplicity
\begin{eqnarray}
<E_1(\tau_1)E_1(\tau_2)>=\int d^3 k \;R + (-\cos^2\alpha) \int
d^3k \; \hat{k}^2_x\; R + (-\sin^2\alpha) \int d^3k \;
\hat{k}^2_y\; R, \nonumber \\
<E_1(\tau_1)E_2(\tau_2)>=<E_2(\tau_1)E_1(\tau_2)>= -\frac{\sin 2
\alpha}{2}\int d^3\;k \; \hat{k}^2_x \;R + \frac{\sin 2
\alpha}{2}\int d^3\;k \; \hat{k}^2_y \;R , \nonumber \\
<E_1(\tau_1)H_3(\tau_2)>=<E_1(\tau_2)H_3(\tau_1)>= -\cos\alpha
\;\int d^3\;k \;\hat{k}_y \; R, \nonumber \\
<E_1(\tau_1)E_2(\tau_2)>=\int d^3\;k\;R + (-\sin^2 \alpha)\int d^3
k \; \hat{k}^2_x\; R + (-\cos^2 \alpha)\int d^3 k \; \hat{k}^2_y\;
R, \nonumber \\
<E_2(\tau_1)H_3(\tau_2)>= <E_2(\tau_2)H_3(\tau_1)>= (-\sin \alpha)
\int d^3 k\; \hat{k}_y \; R, \nonumber \\
<H_3(\tau_1)H_3(\tau_2)>= \int d^3 k \; \hat{k}^2_x \;R+ \int d^3
k \; \hat{k}^2_y \;R,
\end{eqnarray}
and
\begin{eqnarray}
R=h^2_0(\omega) \;\frac{1}{2}\cos k F , \;\;\; F=c \gamma
(\tau_2-\tau_1)[1-\hat{k}_y \frac{v}{c}\frac{\sin
\delta/2}{\delta/2}],\;\;
 \delta=\alpha_2-\alpha_1=\Omega \gamma
 (\tau_2-\tau_1).
\end{eqnarray}
After some  calculations we come to the following expression for
$I_{(11)}$
\begin{eqnarray} \label{eq:CF11}
I_{(11)}=\langle E_{(1)}(\mu_1|\tau_1)E_{(1)}(\mu_2|\tau_2)
\rangle=\gamma^2 \cos \delta \int d^3k \;h^2_0(\omega) \frac{1}{2}
\cos kF + 2 \beta \gamma^2 \cos \frac{\delta}{2} \int d^3 k
\;\hat{k}_y \; h^2_0(\omega)\frac{1}{2} \cos kF + \nonumber
\\
\gamma^2 [\beta^2 - \cos^2 \frac{\delta}{2}]\int d^3 k\;
\hat{k}^2_x \; h^2_0(\omega)\; \cos kF+ \gamma^2[\beta^2 +
\sin^2\frac{\delta}{2}] \int d^3 k\; \hat{k}^2_y
h^2_0(\omega)\frac{1}{2} \cos kF. \nonumber \\
\end{eqnarray}
\indent This function clearly depends only on the proper time
interval $\tau_2-\tau_1$ and does not dependent of the central
proper time $(\tau_1+\tau_2)/2$ that is
\begin{eqnarray}
I_{(11)}=I_{(11)}(\tau_2 - \tau_1).
\end{eqnarray}
 General expressions for other
CFs can be found in  Appendix \ref{sec-generalExpressions}. They
have the same properties and also depend only on the proper time
interval $\tau_2-\tau_1$.
\subsection{
\textbf{The Correlation Function $\langle
E_{(1)}(\mu_1|\tau_1)E_{(1)}(\mu_2|\tau_2) \rangle$: Final
Expression}} \label{sec-Final} \indent The CF  $\langle
E_{(1)}(\mu_1|\tau_1)E_{(1)}(\mu_2|\tau_2) \rangle$ can be
represented in terms of elementary functions .
Completing integration of (\ref{eq:CF11}) over k and then over
$\phi$ we come to the expression :
\begin{eqnarray}
\label{eq-CF11Final}
 \langle
E_{(1)}(\mu_1|\tau_1)E_{(1)}(\mu_1|\tau_2) \rangle = \frac{3\hbar
c}{2 \pi^2 [c(t_2-t_1)]^4} \gamma^2 \{ + [2 \pi \cos \delta]
\int_{0}^{\pi}d\theta\frac{\sin\theta}
{(1-k^2\sin^2\theta)^{7/2}} \nonumber \\
+[3\pi k^2 \cos\delta -2 \pi \cos^2(\delta/2)+ 2\pi \beta^2 - 8
\pi \beta k \cos(\delta/2)+\pi]
\int_{0}^{\pi}d\theta\frac{\sin^3\theta}
{(1-k^2\sin^2\theta)^{7/2}} \nonumber \\
+[-3 \pi k^2 \cos^2(\delta/2) + 3 \pi \beta^2 k^2 -2 \pi \beta k^3
\cos(\delta/2) + 4 \pi k^2]
\int_{0}^{\pi}d\theta\frac{\sin^5\theta}
{(1-k^2\sin^2\theta)^{7/2}} \}
\end{eqnarray}
(see  Appendix \ref{sec-IntegralCalculationFinal} for details).
The integrals over $\theta$ in this expression are :
\begin{eqnarray}
\label{eq-CF11-0Final} \int_{0}^{\pi}d\theta\frac{\sin\theta}
{(1-k^2\sin^2\theta)^{7/2}}= \frac{2}{5(1-k^2)} +
\frac{8}{15(1-k^2)^2} + \frac{16}{15(1-k^2)^3} , \label{eq:sin1} \\
 \int_{0}^{\pi}d\theta\frac{\sin^3\theta}
{(1-k^2\sin^2\theta)^{7/2}}= \frac{4}{15(1-k^2)^2}
+\frac{16}{15(1-k^2)^3}, \label{eq:sin3}\\
\int_{0}^{\pi}d\theta\frac{\sin^5\theta}
{(1-k^2\sin^2\theta)^{7/2}}=\frac{16}{15(1-k^2)^3}
 \label{eq:sin5}
\end{eqnarray}
(see formulas \cite{prudnikov}, 1.5.23, 1.2.43)
and the constant (not a wave vector $\vec{k}$) $k=-\frac{v}{c}\frac{\sin \delta/2}{\delta/2}$.\\
 \\
\indent So the CF $\langle
E_{(1)}(\mu_1|\tau_1)E_{(1)}(\mu_2|\tau_2) \rangle$ can be
represented in terms of elementary  functions. Other CFs  can also
be expressed in  terms of elementary functions.\\
\indent In this form the CFs do not display thermal features. In
the next section we will show that the CFs should be modified to
take their periodicity into consideration. The periodic CF have
thermal features.
\section{The Spectrum of the Random Classical
Electromagnetic Radiation Observed by a Rotating Detector.}
\label{sec-classicalSpectrum}
\subsection{Periodicity of the Correlation Functions. }
\indent The two-field CFs at a rotating detector should be
periodic because CF measurements is one of the tools the detector
can use to justify the periodicity of its motion.  Mathematically
it means that
\begin{eqnarray}
 I_{(11)}(t_2 - t_1) = I_{(11)}((t_2 - t_1)+ \frac{2\pi}
 {\Omega}n )
\end{eqnarray}
or
\begin{eqnarray}
 I_{(11)}(\tau_2 - \tau_1) = I_{(11)}((\tau_2 - \tau_1)+ \frac{2\pi}
 {\Omega \gamma}n )
\end{eqnarray}
  Here
$\Omega=\frac{2\pi}{T}$ is an angular velocity of the rotating
detector and $n=  0, 1, 2, 3,...$. Using (\ref{eq:CF11}) it is
easy to show after some trigonometrical transformations that the
CF is periodic [05.03.06] if in its integrand
\begin{eqnarray}
\omega=\Omega n.
\end{eqnarray}
 It means that the rotating detector observes not entire random
 electromagnetic radiation but only the discrete part of it.
The discrete spectrum  is the same as a rotating electrical charge
radiates \cite{ivanenko}(39.29).\\
\indent The final expression (\ref{eq-CF11Final})  for $I_{(11)}$
can not be used to analyze the periodicity consequences because
the integration over entire continuous spectrum of $\omega$ has
already been done in it.It is why we have used the expression
(\ref{eq:CF11}), before the integration over $\omega$. \\
 \indent Let us now consider the periodic correlation
function $I_{(11)}$ for the discrete spectrum. There are two ways
to do that. The first one is simpler, just to modify the formula
(\ref{eq:CF11}) for $I_{(11)}$ for the discrete spectrum. It will
be described below in the next section. The  second  one is
identical with the approach we have used above for the continuous
spectrum but with the modified equations (\ref{eq:ff}) and
relationships (\ref{eq:thetas}) for the discrete spectrum. It is
described in  Appendix \ref{sec-ModifiedExpressionRandom}.
\subsection{The Correlation Function
 $ \langle
E_{(1)}(\mu_1|\tau_1)E_{(1)}(\mu_2|\tau_2) \rangle $ \textbf{with
the Discrete Spectrum: General Expression} }
\label{sec:CFDiscrete} The integrands in (\ref{eq:CF11}) can be
changed according to an obvious relationship
\begin{eqnarray}
\label{eq:CF_Integral} \int d^3 k [ \;\; ] \frac{1}{2} h^2_0
(\omega) \cos kF =
 \frac{c \hbar k^4_0}{4 \pi^2}\int d O [\;\;] S,
\end{eqnarray}
where
\begin{eqnarray}
S=\int d \kappa \;\kappa^3 \cos \kappa F_d , \;\;\; d O =d\theta
d\phi \sin \theta, \;\;\; \kappa=\frac{k}{k_0},
\end{eqnarray}
and
\begin{eqnarray}
 F_d=k_0 F= \delta [1 - \hat{k}_y \frac{v}{c}\frac{\sin \delta /2}
 {\delta/2}],
\end{eqnarray}
The expressions in $[\;\;\;]$ are 1, $\hat{k}_y$, $\hat{k}_x^2$,
or $\hat{k}_y^2$ and  do not depend on
$\kappa=k / k_0$. \\
\indent For the discrete spectrum case the integration in
(\ref{eq:CF_Integral}) over $\kappa$ should be changed to
summation over n. So the the only term to be changed is S. It
becomes ( in new notations)
\begin{eqnarray}
\label{eq:CF11d2} S_d=\sum_{0}^{\infty} n^3 \cos n F_d.
\end{eqnarray}
Then the periodical CF, corresponding to (\ref{eq:CF11}),  with
the discrete spectrum  can be given in the form
\begin{eqnarray}
\label{eq:CF11d1}
 \langle
E_{(1)}(\mu_1|\tau_1)E_{(1)}(\mu_2|\tau_2) \rangle_d=  \frac{c
\hbar k^4_0}{4 \pi^2} \;\; \{ \;\;\gamma^2 \cos \delta \int d O \;
S_d+ 2 \beta \gamma^2 \cos \frac{\delta}{2} \int dO \;\hat{k}_y
\;S_d
 + \nonumber \\
\gamma^2 [\beta^2 - \cos^2 \frac{\delta}{2}]\int d O\; \hat{k}^2_x
\; S_d+ \gamma^2[\beta^2 + \sin^2\frac{\delta}{2}] \int d O\;
\hat{k}^2_y S_d \;\;\}.
\end{eqnarray}
\indent The physical sense of the correlation functions with the
discrete spectrum  can be understood if we use the Abel-Plana
formula to analyze $S_d$. It will be done in next sections.
\subsection{ The Abel-Plana Formula. }
\indent The Abel-Plana summation formula
\cite{Bateman1953},\cite{MT1988}, \cite{Evgrafov1968}has the form:
\begin{eqnarray}
\label{eq:AbelPlana}
 \sum_{n=0}^\infty \, f(n)= \int_0^\infty
f(x)\,dx + \frac{f(0)}{2} +i \,\int_0^\infty \,dt \,
\frac{f(it)-f(-it)}{e^{2 \pi t}-1},
\end{eqnarray}
 Having utilized this formula for (\ref{eq:CF11d2})
with $f(n)= n^3 \cos n F_d$ and following  \cite{MT1988} we come
to the following expression
for $S_d$
\begin{eqnarray}
\label{eq:sum}
 \Omega^4 S_d
 =  \int _0^{\infty} d \, \omega \omega^3 \cos( \omega \tilde{F})+
 \int_0^{\infty} d\omega \frac{2 \omega^3 \cosh(\omega \tilde{F})}{e^
 {2\pi \omega/\Omega}-1}, & \tilde{F}=\frac{F_d}{\Omega},
\end{eqnarray}
or after integration to:
\begin{eqnarray}
\label{eq:I_one}
 S_d = \frac{6}{F_d^4} + [\frac{3 -2 \sin^2(F_d/2)}{8
\sin^4(F_d/2)} -\frac{6}{F_d^4}].
\end{eqnarray}
 So now
$S_d$ is broken into divergent and convergent parts and  both
expressions are very similar to \cite{boyer1980} (74). (Another
form
of $S_d$ is given in  Appendix \ref{sec-anotherExpressionSd}.)
\\
\indent Let us compare the expression (\ref{eq:sum}) for
$\emph{S}_d$ with the expression \cite{boyer1980} (74)
\begin{eqnarray}
\label{eq:Tspectral function} \int_0^{\infty} d\omega \omega^3
\coth(\frac{\hbar \omega}{2 k T})\cos \omega t=
\int_0^{\infty}d\omega\omega^3 \cos \omega t + \int_0^{\infty}d
\omega \frac{2 \omega ^3}{e^{\frac{\hbar \omega}{k T}-1}} \cos
\omega t
\end{eqnarray}
which is the Fourie component of the spectral function
$\frac{1}{2}\hbar \omega \coth \frac{\hbar \omega}{2 k T}$ of the
electromagnetic radiation with Planck's spectrum at the
temperature T, with  the zero-point radiation included.\\
 The right sides of these expressions are very similar
and have the Planck's factor $1/(\exp^{\frac{\hbar \omega}{k_B}T}-
1 )$ if
 we define in (\ref{eq:sum}) a new
constant, a rotation temperature,  $T_{rot}$ according to
\begin{equation}
\label{eq:T}
 T_{rot}=\frac{\hbar \Omega}{2 \pi k_B}.
\end{equation}
The Planck's factor is a sign that some thermal effect accompany
the detector rotation in the random classical zero-point
electromagnetic radiation. There is also a significant distinction
between them though. In the first expression
\emph{$\tilde{F}=t(1-\hat{k}_y)\frac{v}{c}\frac{\sin\delta/2}{\delta/2}$}
and \emph{cosh} are used instead of \emph{t} and \emph{cos}
respectively in (\ref{eq:Tspectral function}).  The coefficient
$F_d$ depends on both $\theta$ and $\phi$ because $\hat{k}_y=\sin
\theta \sin\phi$. It means that the thermal radiation observed by
the rotating detector defers from the Planck's radiation and it is
anisotropic.\\
 \indent For $\tilde{F}=0$ and t=0, when two observation points,
 $\tau_1$ and $\tau_2$,
 coincide  the right sides of both
 expressions are identical.This remarkable resemblance brings up the
idea that the energy density, one-observation-point quantity, of
the random classical electromagnetic radiation measured by a
detector, rotating through a zero point radiation, has the Planck
spectrum at the temperature $T_{rot}$ (\ref{eq:T}). This issue
will be discussed in the next section.

\subsection{  The Energy Density of Random Classical
Electromagnetic Radiation Observed by a Rotating detector and
Planck's Spectrum} \indent Using (\ref{eq:FieldAtTetrad}) the
energy density measured by the rotating observer at a reference
frame $\mu$
\begin{equation}\label{eq:energy density}
w(\mu)=\frac{1}{8 \pi} \sum_{a=1}^3\;( \;\;\langle
E_{(a)}^2(\mu|\tau)\rangle + \langle H_{(a)}^2(\mu|\tau)\rangle
\;\; )
\end{equation}
can be given in terms of electric and magnetic fields measured in
the laboratory coordinate system
\begin{equation}\label{eq:energy density}
w(\mu)=\frac{1}{4\pi}\{\;\;[\;\langle E_1^2 \rangle + \langle
E_3^2 \rangle \; ]\gamma^2(1+\beta^2)+\langle E_2^2 \rangle
\;\;\},
\end{equation}
where as we will show below $\langle E_i^2 \rangle =\langle H_i^2
\rangle $, $i=1,2,3
$. \\
\indent With the help of the formulas from  Appendix
\ref{sec-ModifiedExpressionRandom} and using the technique,
described above for a discrete spectrum we come to the following
expressions
\begin{eqnarray}
\label{eq:FiledSquared} \langle E_i^2 \rangle = \langle H_i^2
\rangle = \frac{k_0^4 \hbar c}{2 \pi^2} \int dO (1-\hat {k}^2_i
)\sum^\infty_{n=0}n^3
\end{eqnarray}
for $i=1,2,3$ and finally after integration over $\theta$ and
$\phi $ :
\begin{equation}
\label{eq: energy density 1} w(\mu)= \frac{ (4\gamma^2-1)}{3
}\;\frac{\hbar}{c^3 \pi^2}\; \Omega^4\sum_{n=0}^\infty n^3
\end{equation}
If we take into consideration the Abel-Plana formula
(\ref{eq:sum}) for $F_d=0$ this expression can be given in a form
more convenient  for physical interpretation
\begin{equation}
\label{eq:energy density 2} w(\mu)= \frac{2 \;(4\gamma^2-1)}{3 }
\; w_{em}(T_{rot}),
\end{equation}
where
\begin{eqnarray}
\label{eq: energy density 3} w_{em}(T_{rot})=\frac{\hbar }{c^3
\pi^2} \;( \; \int _0^{\infty} d \, \omega \frac{1}{2}\omega^3 +
 \int_0^{\infty} d\omega \frac{ \omega^3 }
 {e^{\hbar\omega/kT_{rot}} -1}
\;).
\end{eqnarray}
\indent
  $w_{em}(T)$ is  the  energy
  density of the electromagnetic field  observed by an inertial
 detector in Planck's spectrum at  the temperature
 T  \cite{boyer1980} . It consists of two
parts. The first one, divergent, is a zero-point energy density
and the second one, convergent, is a part of the energy density
due to a temperature $T_{rot}$ \cite{Milonni1994}.  \\
\indent Thus the detector rotating in the zero-point  radiation
under the temperature $T=0$  observes not only  original
zero-point radiation but also the radiation determined by the
parameter $T_{rot}$. We can interpret $T_{rot}$ now as the
temperature associated with the detector rotation.
\\
\indent Indeed, after integration over $\omega$ the second
convergent term is
\begin{eqnarray}
\label{eq:mainFormula}
 w_T(\mu) = \frac{2(4\gamma^2-1)}{3}\; w_{black},
\end{eqnarray}
where
\begin{eqnarray}
w_{black} = 4 \; \frac{\pi^2 k_B^4}{60 (c \hbar)^3}\; T_{rot}^4=
\frac{4 \sigma}{c} T_{rot}^4
\end{eqnarray}
is the energy density of the black radiation at the temperature
$T_{rot}$ [\cite{landau_lifshits}, (60,14)],
 $k_B$ is the Boltzman
constant, and $\sigma$ is the Stefan-Boltzman constant.\\
\indent So, due to periodicity of the motion, the energy density
observed by a detector rotating through a zero-point radiation and
the energy density observed an inertial detector in a thermal bath
at the temperature $T_{rot}=\frac{\hbar \Omega}{2 \pi k}$ are
connected by the formula (\ref{eq:mainFormula}). The factor
$\frac{2}{3}(4\gamma^2-1)$ comes from integration in
(\ref{eq:FiledSquared}) over angles and  is a consequence of
anisotropy of the electromagnetic field measured by the rotating
observer.
\section{Classical Massless Zero-Point Scalar Field at
Rotating Detector. } \label{sec-massless}
\subsection{ Correlation Function  and Tetrads.} \indent The field $\psi_s(\mu_{\tau}|\tau)$ in a
tetrad $\mu_\tau$
 has the same form as in the laboratory
coordinate system, $\psi_s(\tau)$ , taken in the location of the
tetrad, because it is a scalar. Then the correlation function
measured by an observer rotating through a classical massless
zero-point scalar field radiation has the form:
\begin{eqnarray}
\langle \psi_s(\mu_1|\tau_1) \psi_s(\mu_2 |\tau_2)\rangle =
\langle \psi_s(\tau_1) \psi_s(\tau_2) \rangle ,
\label{eq:CF_Scal_Field_Continuous}
\end{eqnarray}
where \cite{boyer1980}
\begin{eqnarray}
\label{eq:scalarField} \psi_s(\tau_i)= \int d^3k_i f(\omega_i)
\cos \{ \vec{k}_i \vec{r}(\tau_i) -\omega_i \gamma \tau_i -
\theta(k_i)\}, \;\; i=1, 2
\end{eqnarray}
and
\begin{eqnarray}
\label{eq:scalarTheta}
 \langle \cos \theta(\vec{k}_1) \cos \theta (\vec {k}_2) \rangle
= \langle \sin \theta(\vec{k}_1) \sin \theta (\vec {k}_2) \rangle
= \frac{1}{2} \delta^3(\vec{k}_1 - \vec{k}_2), \;\;
f^2(\omega)=\frac{\hbar c^2}{2 \pi^2 \omega} \nonumber \\
 \langle \cos \theta(\vec{k}_1) \sin \theta (\vec {k}_2) \rangle
 =0,
  \;\;\;\;  \vec{r}(\tau_i)=(a \; \cos \Omega \gamma \tau_i,\;
  a \;\sin \Omega
\tau_1, \; 0 ).
\end{eqnarray}
Using these expressions and substitution (\ref{eq:VariableChange})
in the integrand we get the expression :
\begin{eqnarray}
\label{eq:ScalarCF}
 \langle \psi_s(\mu_1 |\tau_1) \psi_s(\mu_2 |\tau_2)\rangle=
 \int
d^3 k f^2(\omega)\frac{1}{2} \cos kF.
\end{eqnarray}
After integration over positive $k= \omega /c$, it becomes:
\begin{eqnarray}
\langle \psi_s(\mu_1 |\tau_1) \psi_s(\mu_2
|\tau_2)\rangle=-\frac{\hbar c}{4\pi^2}\:\int_0^\pi d\theta
\:\sin\theta \: \int_0^{2 \pi} d\phi \: \: [E \:\sin \phi-
B]^{-2},
\end{eqnarray}
where $B=\gamma \tau c$, $E=2 a \sin\theta \: \sin \frac{ \Omega
\gamma \tau}{2}$, and $\tau=\tau_2- \tau_1$. \\
Because $B-|E|= c \gamma \tau \{1-\frac{v}{c} |\sin \theta \:
\frac{\sin \pi (\gamma \tau /T)}{\pi(\gamma \tau / T)}|\}> c
\gamma \tau (1 - v/c) >0,$
 and using \cite{prudnikov}
 we obtain :
\begin{eqnarray}
\int_0^{2 \pi} d\phi \: \frac {1}{[E \:\sin \phi- B]^2}= \frac{2
\pi B}{(B^2 - E^2)^{3/2}}.
\end{eqnarray}
 Having integrated over $\theta$ we come to the final expression
 for the CF of the random classical
 massless
 scalar field at the rotating  detector moving through a zero point
 fluctuating massless scalar radiation:
\begin{eqnarray}
\langle \psi_s(\mu_1|\tau_1) \psi_s(\mu_2 |\tau_2)\rangle=
  -\frac{\hbar
c}{\pi}\frac{1}{(\gamma (\tau_2 - \tau_1)c)^2- 4 r^2 \sin^2
\frac{\Omega \gamma (\tau_2 - \tau_1)}{2}}.
\end{eqnarray}
\indent This correlation function received in the classical
approach  is identical to the positive frequency Wightman function
\cite{davis1996} (3) up to a constant. \\
 \indent In the scalar field CF is periodical for the same reasons
 it is periodical in the electromagnetical fields.
We have not yet taken into consideration the periodicity of this
CF connected with  the rotation of the detector in the scalar
field. This issue is investigated below.\\
\subsection{Periodicity of the Correlation Function, Abel-Plana
formula, and the Planck's Factor.}
 \indent  The equation
(\ref{eq:ScalarCF}) can be given in the form
\begin{eqnarray}
 \langle \psi_s(\mu_1 |\tau_1) \psi_s(\mu_2|\tau_2)\rangle=
 \frac{\hbar c k^2_0}{4 \pi ^2}\int
d O \int d \kappa \kappa \cos \kappa F_d.
\end{eqnarray}
If this function is periodic its spectrum should be $\kappa
=\frac{c k}{c k_0}=n=0,1,2,..$ and then
\begin{eqnarray}
\label{eq:ScalerCFd}
 \langle \psi_s(\mu_1 |\tau_1) \psi_s(\mu_2 |\tau_2)\rangle_d=
 \frac{\hbar c k^2_0}{4 \pi ^2}\int
d O \sum_{n=0}^{\infty} n \cos n F_d.
\end{eqnarray}
 \indent Abel-Plana summation formula
in this case is
\begin{eqnarray}
\sum_{n=0}^{\infty}n \cos nF_d =\int_0^{\infty} dt\; t \cos tF_d -
\int_0^{\infty} dt \frac{2t \cosh tF_d }{e^{2 \pi t}-1}
\end{eqnarray}
or
\begin{eqnarray}
\label{eq:planck} \Omega^2 \;\sum_{n=0}^{\infty}n \cos nF_d
=\int_0^{\infty} d\omega\; \omega \cos \omega \tilde{F} -
\int_0^{\infty} d\omega \frac{2\omega \cosh \omega \tilde{F}
}{e^{\frac{\hbar \omega}{k T_{rot}}}-1},
\end{eqnarray}
where $T_{rot}$ is defined in (\ref{eq:T}). \\
Then
\begin{eqnarray}
\langle \psi_s(\mu_1 |\tau_1) \psi_s(\mu_2 |\tau_2)\rangle_d=
 \frac{\hbar}{4 \pi ^2 c}\int d O \; \{ \;\int_0^{\infty} d\omega\; \omega \cos \omega \tilde{F} -
\int_0^{\infty} d\omega \frac{2\omega \cosh \omega \tilde{F}
}{e^{\frac{\hbar \omega}{k T_{rot}}}-1}\; \}
\end{eqnarray}
 \indent The  expression in $\{ \;\; \}$ is similar to the
right side of the  expression \cite{boyer1980}, (27) for the
correlation function of the detector at rest in Planck's spectrum
at the temperature T
\begin{eqnarray}
\int_0^{\infty}d \omega \omega \coth\frac{\hbar \omega}{2
kT}\cos\omega t =\int_0^{\infty}d \omega \cos \omega t +
\int_0^{\infty}d \omega \frac{2 \omega \cos\omega
t}{e^{\frac{\hbar \omega}{kT}}-1}.
\end{eqnarray}
 The appearance of the Planck's
factor $(e^{\frac{\hbar \omega}{kT_{rot}}}-1)^{-1}$  shows the
similarity between the radiation observed at the rotating detector
in the massless scalar zero-point field  and the radiation
spectrum observed by an inertial observer placed in a thermostat
filled up with the radiation at the temperature $T=T_{rot}$. But
there is also distinction between them. F and $\cosh$ are used in
the first expression whereas t and $\cos$ are used in the second
expression respectively.
 The $\tilde{F}$ is a function of $\theta$ and $\phi$. It means that a
thermal radiation observed by the rotating detector moving in the
massless scalar
zero-point radiation is anisotropic.  \\
 \indent  The  resemblance between both expressions  becomes closer
 if $t=0$ and
$\tilde{F}=0$ and two points of an observation agree. Both
expressions are identical. But in the case of one-point
observation which occurs when $\tilde{F}=0$ it is better to
consider the energy density of the scalar massless field. This
brings us to the next section.
\subsection{The Energy Density of Random Classical Massless Scalar
Field Observed by a Rotating Detector and Planck's Spectrum.}
\indent Let us consider the energy density $\langle
T_{(44)}(\mu)\rangle$ of the massless scalar field at  the
detector rotating through the zero-point massless scalar field. It
can be expressed in terms of the tensor of energy-momentum
$T_{ik}$ at the location of the detector in the laboratory
coordinate system \cite{synge} as
\begin{eqnarray}
\langle T_{(44)}(\mu)\rangle=\mu_{(4)}^i\mu_{(4)}^k \langle
T_{ik}, \rangle
\end{eqnarray}
where $\mu_{a}^i$ are  tetrads. The  energy-momentum tensor is
\cite{birell1982}(2.27)
\begin{eqnarray}
T_{ik}= \psi_{,i}\psi_{,k}
-\frac{1}{2}\eta_{ik}\eta^{rs}\psi_{,r}\psi_{, s}, \;\; \eta_{ik}=
\eta_{ik}=diag(1,1,1,-1)
\end{eqnarray}
Using (\ref{eq:scalarField}), (\ref{eq:scalarTheta}), and
Frenet-Serret tetrads it is easy to show that
\begin{eqnarray}
\langle T_{11}\rangle=\langle T_{22}\rangle=\langle
T_{33}\rangle=\frac{1}{3}\langle T_{44}\rangle = \frac{\hbar
c}{\pi} \int dk k^3 =\frac{\hbar \Omega^4}{\pi c^3} \int d\kappa
\;\; \kappa^3
\end{eqnarray}
and
\begin{eqnarray}
\langle T_{(44)}(\mu)\rangle = \frac{4 \gamma^2 -1}{3} \langle
T_{44} \rangle = \frac{4 \gamma^2 -1}{3}  \frac{\hbar
\Omega^4}{\pi c^3} \int d\kappa \;\; \kappa^3
\end{eqnarray}
This expression with periodical features taken into consideration
has the following form
\begin{eqnarray}
\langle T_{(44)}(\mu)\rangle_d  = \frac{4 \gamma^2
-1}{3}\;\;\frac{\hbar}{\pi c^3}\;  \Omega^4\; \sum_{n=0}^{\infty}
n^3
\end{eqnarray}
or
\begin{eqnarray}
\langle T_{(44)}(\mu)\rangle_d =\frac{4 \gamma^2
-1}{3}\;\;\frac{\hbar}{\pi c^3}\;2 \;( \; \int _0^{\infty} d \,
\omega \frac{1}{2}\omega^3 +
 \int_0^{\infty} d\omega \frac{ \omega^3 }
 {e^{\hbar\omega/kT_{rot}} -1}.
\;)
\end{eqnarray}
\indent Let us compare this expression and the expression for the
energy density of the massless scalar field  with the Planck's
spectrum of random thermal radiation at the temperature T along
with the zero-point radiation in an inertial reference frame
\begin{eqnarray}
\langle T_{44} \rangle_T=\frac{1}{2} [(\frac{\partial
\psi_T}{\partial (ct)})^2 + (\frac{\partial \psi_T}{\partial x
})^2 + (\frac{\partial \psi_T}{\partial y})^2 + (\frac{\partial
\psi_T}{\partial z})^2],
\end{eqnarray}
where \cite{boyer1980}
\begin{eqnarray}
\psi_T =\int d^3k\; f_T (\omega)\; \cos\;[\;\vec{k}\vec{r} -
\omega t -\theta(\vec{k})\;]
\end{eqnarray}
and
\begin{eqnarray}
f^2_T(\omega)=\frac{c^2}{\pi^2}\; \frac{\hbar}{\omega}\; [\;
\frac{1}{2} + \frac{1}{\exp(\hbar \omega / k T) -1 }\;].
\end{eqnarray}
It is easy to show  that
\begin{eqnarray}
\langle T_{44} \rangle_T= \frac{3 \hbar}{\pi c^3}( \; \int
_0^{\infty} d \, \omega \frac{1}{2}\omega^3 +
 \int_0^{\infty} d\omega \frac{ \omega^3 }
 {e^{\hbar\omega/kT} -1}),
\end{eqnarray}
and
\begin{eqnarray}
\langle T_{(44)}(\mu) \rangle_d= \frac{2 (4 \gamma^2 -1)}{9}
\langle T_{44} \rangle_{T_{rot}}.
\end{eqnarray}
\indent So, due to periodicity of the motion, an observer rotating
through a zero point radiation of a massless random scalar field
should see the same energy density as would see  an inertial
observer moving in a thermal bath at the temperature
$T_{rot}=\frac{\hbar \Omega}{2 \pi k}$ when multiplied by the
factor $\frac{2}{9}(4\gamma^2-1)$. This factor comes from
integration over angles and  is a consequence of anisotropy of the
scalar field measured by an observer with velocity $\beta$.
\section{Discussion}
\indent The thermal effects of non inertial motion investigated in
the past for uniform acceleration through classical random
zero-point radiation of electromagnetic and massless scalar field
are shown to exist for the case of
rotary motion also. \\
\indent The rotating reference system $\{\mu_\tau\}$, along with
the two-point correlation  functions (CFs) and energy density, is
defined and used as the basis for investigating  effects observed
by  a detector rotating through random classical zero-point
radiation. The reference system consists of Frenet -Serret
orthogonal tetrads
 $\mu_\tau$. At each proper time $\tau$ the rotating detector is
 at rest  and has a constant acceleration vector at the
 $\mu_\tau$.
 \\
\indent The two-point CFs and the energy density at the rotating
reference system should be periodic with the period $T=\frac{2
\pi}{\Omega}$, where $\Omega$ is an angular detector velocity,
because CF and energy density measurements is one of the tools the
detector can use to justify the periodicity of its motion. The CFs
 have been calculated  for both \emph{electromagnetic} and
 \emph{massless scalar} fields in two cases, with and without taking
  this periodicity into
consideration. It turned out that only periodic CFs have some
thermal features and  particularly the Planck's factor with the
temperature $T_{rot}=\frac{\hbar \Omega}{2 \pi k_B}$ ($k_B$ is the
Boltzman constant). Mathematically this property is connected with
the discrete spectrum of the periodic CFs  and its interpretation
based on the Abel-Plana
summation formula. \\
\indent It is also shown that energy densities of the
electromagnetic and massless scalar fields observed by the
rotating detector are respectively
\begin{eqnarray} \nonumber
w(\mu)= \frac{2 \;(4\gamma^2-1)}{3 }\; w_{em}(T_{rot})
\end{eqnarray}
and
\begin{eqnarray} \nonumber
\langle T_{(44)}(\mu) \rangle_d= \frac{2 (4 \gamma^2 -1)}{9}
\langle T_{44} \rangle_{T_{rot}}.
\end{eqnarray}
The  $w_{em}(T_{rot})$ and $\langle T_{44} \rangle_{T_{rot}}$ are
Planck's energy densities of electromagnetic field and massless
scalar field respectively observed by an inertial detector at the
temperature $T_{rot}$ along with their random zero-point
radiation. \\
\indent This thermal effect is masked by  factor
$\frac{2}{3}(4\gamma^2-1)$ for the electromagnetic field and
$\frac{2}{9}(4\gamma^2-1)$ for the massless scalar field, where
$\gamma=(1-(\frac{\Omega r}{c})^2)^{-1/2}$. Appearance of these
masking factors is connected with the fact that rotation is
defined by two parameters, angular velocity and the radius of
rotation, in contrast with a uniformly accelerated linear motion
which is defined by only one parameter, acceleration a. As a
consequence the thermal effects observed by detectors rotating
with the same angular velocity on different circumferences are
different. The further detector is from the rotation center the
greater energy density it sees. If $r \rightarrow \infty$ then
$w(\mu)\rightarrow \infty$ and $\langle T_{(44)}(\mu) \rangle_d
\rightarrow \infty$. Appearance of the masking factors is
connected with the fact that rotation is defined by two
parameters, angular velocity and the radius of rotation, in
contrast with a uniform accelerated  linear motion. The latter is
defined by only one parameter, acceleration "a". \\
\indent Some of the results discussed in this paper have been
obtained in \cite{levin2006} using reference systems consisting of
{\textbf{global}} instantaneous inertial reference frames, not
tetrads.\\

{\bf APPENDIX} \\
\appendix
\section{ Frenet-Serret Orthogonal Tetrads}
\label{sec:FrenetSerret}
 \indent The Frenet-Serret orthogonal
tetrad mentioned in the section \ref{sec-Definition} can be
associated with each point of the time-like world line of a moving
detector. The vectors of the tetrad are defined by the equations
\cite{synge}(55):
\begin{eqnarray}
D\mu^i_{(4)}=b\mu^i_{(1)}, \nonumber \\
D\mu^i_{(1)}= \tilde{c}\mu^i_{(2)} + b \mu^i_{(4)}, \nonumber \\
D\mu^i_{(2)}=d \mu^i_{(3)} -\tilde{c} \mu^i_{(1)}, \nonumber \\
D\mu^i_{(3)}=-d \mu^i_{(2)}
\end{eqnarray}
together with
\begin{eqnarray}
\mu^i_{(4)}\mu_{{(4)}i}=-1, \;
\mu^i_{(1)}\mu_{{(1)}i}=\mu^i_{(2)}\mu_{{(2)}i}=
\mu^i_{(3)}\mu_{{(3)}i}=1.
\end{eqnarray}
They  are also orthogonal :
\begin{eqnarray}
\mu^i_{(a)} \mu_{{(b)}i}=\eta_{(ab)}, \;\;
\eta_{(ab)}=diag(1,1,1,-1), \;\; a,b =1,2,3,4.
\end{eqnarray}
 In the flat space-time,  with metrics $g_{ik}$=diag(1,1,1,-1),
  $D=\frac{d}{d \tau}$ where $\tau$
is a proper time of the detector. \\
\indent Let
 \begin{eqnarray}
 \mu^i_{(4)}= \frac{U^i}{c}= \mu^i_{(4)}=(-\beta \gamma \sin \alpha,\beta \gamma \cos \alpha,
0, \gamma),
 \end{eqnarray}
 where $U^i$ is a 4 vector velocity of the rotating detector
 in the laboratory coordinate system, $\beta=
 v/c=\Omega a /c, \gamma=(1-\beta^2)^{-1/2}, \alpha= \Omega
 \gamma \tau$, and $\Omega, \;a$ are an angular velocity and
 the circumference radius respectively of the rotating detector.
 It is easy to check that the set of the  4-vectors
 (\ref{eq:FStetrad})
  is a solution to this equation system, with the coefficients
\begin{eqnarray}
 b=-\beta \Omega \gamma^2, \; \tilde{c}=
 \Omega \gamma^2, ;\ d=0.
\end{eqnarray}
In a Fermi-Walker tetrad \cite{moller72} (9.148, 4.139, and
4.167),
tetrad vectors are defined as
\begin{eqnarray}
\frac{d e_{(a)k}}{d \tau}= (e_{(a)l}\dot{U}_l)U_k/c^2
-(e_{(a)l}U_l)\dot{U}_k/c^2
\end{eqnarray}
and can be given in the form \cite{moller72} 4.167
\begin{eqnarray}
e_{(1)k}= (\cos\alpha \cos \alpha \gamma + \gamma \sin\alpha
\sin\alpha\gamma,\; \sin\alpha \cos \alpha \gamma - \gamma \cos
\alpha \sin \alpha \gamma,0,-i(v \gamma/c)\sin \alpha \gamma),
\nonumber \\
e_{(2)k}= (\cos\alpha \sin \alpha \gamma - \gamma \sin\alpha
\cos\alpha\gamma,\; \sin\alpha \sin \alpha \gamma + \gamma \cos
\alpha \cos \alpha \gamma,0,+ i(v \gamma/c)\cos \alpha \gamma),
\nonumber \\
e_{(3)k}=(0, 0, 1, 0), \nonumber \\
e_{(4)k}= (i \frac{v}{c}\gamma \sin \alpha, -i\frac{v}{c}\gamma
\cos \alpha ,0 ,\gamma).
\end{eqnarray}
In \cite{moller72} 4.167  the metrics is chosen in the form
$g_{ik}=(1,1,1,1)$.
\section{ Other Correlation
Functions of Electromagnetic Field at a Rotating Detector:General
Expressions} \label{sec-generalExpressions}
 \indent The general expression for $\langle
E_{(1)}(\mu_1|\tau_1)E_{(1)}(\mu_2|\tau_2) \rangle$ has been
received in the
 section (\ref{sec-general})
 The other diagonal electric field components of the CFs are as
 follows:\\
\begin{eqnarray}
\langle E_{(2)}(\mu_1 | \tau_1)\:E_{(2)}(\mu_2| \tau_2)\rangle=
\langle E_1(\tau_1)E_1(\tau_2)\rangle \sin \alpha_1 \sin \alpha_2
+ \langle E_1(\tau_1)E_2(\tau_2)\rangle
(-1)\sin \alpha_1 \cos \alpha_2 + \nonumber \\
\langle
E_2(\tau_1)E_1(\tau_2)\rangle (-1)\cos\alpha_1 \sin \alpha_2 +
\langle E_2(\tau_1)E_1(\tau_2)\rangle \cos \alpha_1 \cos \alpha_2
\end{eqnarray}
 \\
\begin{eqnarray}
\langle E_{(3)}(\mu_1 | \tau_1)\:E_{(3)}(\mu_2| \tau_2)\rangle=
-\gamma^2\frac{v}{c} \cos \alpha_1 \langle
E_3(\tau_2)H_1(\tau_1)\rangle - \gamma^2\frac{v}{c} \cos \alpha_2
\langle H_1(\tau_2)E_3(\tau_1)\rangle - \nonumber
\\
\gamma^2\frac{v}{c} \sin \alpha_1 \langle
E_3(\tau_2)H_2(\tau_1)\rangle- \gamma^2\frac{v}{c} \sin \alpha_2
\langle H_2(\tau_2)E_3(\tau_1)\rangle+ \gamma^2(\frac{v}{c})^2
\cos \alpha_2 \cos \alpha_1  \langle H_1(\tau_2)H_1(\tau_1)\rangle
+ \nonumber \\
\gamma^2(\frac{v}{c})^2 \sin \alpha_2 \sin \alpha_1 \langle
H_2(\tau_2)H_2(\tau_1)\rangle + \gamma^2(\frac{v}{c})^2 \cos
\alpha_2 \sin \alpha_1 \langle H_1(\tau_2)H_2(\tau_1)\rangle +
\nonumber \\
\gamma^2(\frac{v}{c})^2 \sin \alpha_2 \cos \alpha_1 \langle
H_2(\tau_2)H_1(\tau_1)\rangle
\end{eqnarray}
Is easy to show that they  depend on the difference
$\delta=\alpha_2 -\alpha_1$ only. \\
\begin{eqnarray}
\langle E_{(2)}(\mu_1 | \tau_1)\:E_{(2)}(\mu_2| \tau_2)\rangle=
\cos\delta \int d^3 k \;R + \sin^2 \frac{\delta}{2}\int d^3 k
\hat{k}^2_x \;R + (-1)\cos^2\frac{\delta}{2} \int d^3 k
\hat{k}^2_y \;R.
\end{eqnarray}
\begin{eqnarray}
\langle E_{(3)}(\mu_1 | \tau_1)\:E_{(3)}(\mu_2| \tau_2)\rangle=
\gamma^2 \frac{v^2}{c^2} \cos \delta\;\int d^3 k\;R + \gamma^2
\frac{v}{c}(-2)\cos\frac{\delta}{2} \int d^3 k\;\hat{k}_y \;R+
\nonumber \\
\gamma^2[1-\frac{v^2}{c^2}\cos^2 \frac{\delta}{2}]\;\int d^3 k
\hat{k}^2_x \;R +
\gamma^2[1+\frac{v^2}{c^2}\sin^2\frac{\delta}{2}]\int d^3
k\;\hat{k}^2_y\;R
\end{eqnarray}
\indent The non diagonal components of the correlation function
are zeroes :
\begin{eqnarray}
\langle E_{(1)}(\mu_1 | \tau_1)\:E_{(2)}(\mu_2| \tau_2)\rangle=
\langle E_{(1)}(\mu_2 | \tau_2)\:E_{(2)}(\mu_1| \tau_1)\rangle=0,
\nonumber \\
\langle E_{(1)}(\mu_1 | \tau_1)\:E_{(3)}(\mu_2| \tau_2)\rangle=
\langle E_{(1)}(\mu_2 | \tau_2)\:E_{(3)}(\mu_1| \tau_1)\rangle=0,
\nonumber \\
\langle E_{(2)}(\mu_1 | \tau_1)\:E_{(3)}(\mu_2| \tau_2)\rangle=
\langle E_{(2)}(\mu_2 | \tau_2)\:E_{(3)}(\mu_1| \tau_1)\rangle=0,
\end{eqnarray}
Similar expressions have been received for the CF with magnetic
field components. So all CFs can be given as 3-dimensional
integrals over $(k, \theta, \phi)$.
\section{Integral calculations:
final expression for $\langle
E_{(1)}(\mu_1|\tau_1)E_{(1)}(\mu_2|\tau_2) \rangle$.}
\label{sec-IntegralCalculationFinal} \indent All non zero
expressions for CFs in the section (\ref{sec-Final}) should be
integrated over $k, \theta$ , and $\phi$. The integral over $k$
can be easily calculated:
\begin{eqnarray}
\label{eq:IntegralOverK}
 \int_0^\infty d k k^3 \cos \{ k ( 2 r \sin
 \frac{\delta}{2} \sin \theta \sin \phi -c (t_2 -t_1))\} =  \frac
{6}{ {\{2 r \sin \frac{\delta}{2}  \sin \theta \sin \phi -
c(t_2-t_1)\}^4}}= \nonumber \\
= \frac {6}{[ c(t_2-t_1)]^4} \frac{1}{ [1-\frac{v}{c}\frac{\sin
\delta/2}{\delta/2}\sin\theta \sin \phi]^4 }.
\end{eqnarray}
The integrals over $\theta$ and $\phi$ can be represented  in
terms of elementary functions. Let us show it  for $\langle
E_{(1)}(\mu_1|\tau_1)E_{(1)}(\mu_1|\tau_2) \rangle$
\begin{eqnarray}
\langle E_{(1)}(\mu_1|\tau_1)E_{(1)}(\mu_1|\tau_2) \rangle =
  \frac{3\hbar c}{2 \pi^2
[c(t_2-t_1)]^4} \gamma^2 \int_{0}^{\pi} d\theta  \nonumber \\
\times \{ \;\; (\cos \delta \sin \theta +(-\cos^2
\frac{\delta}{2}+ \frac{v^2}{c^2}) \sin^3 \theta) \int_{0}^{2\pi}
d\phi \:
\frac{1}{(1+b \sin\phi)^4} \nonumber \\
+ (-2\frac{v}{c}\cos\frac{\delta}{2})sin^2\theta\int_{0}^{2\pi}
d\phi \: \frac{\sin\phi}{(1+b \sin\phi)^4}
+\sin^3\theta\int_{0}^{2\pi} d\phi \: \frac{\sin^2\phi}{(1+b
\sin\phi)^4} \;\; \},
\end{eqnarray}
We have taken into consideration here that
\begin{eqnarray}
\label{eq:HatVector} \hat{k}_x = \sin\theta \cos \phi ,  &
\hat{k}_y= \sin\theta \sin\phi ,  & \hat{k}_z= \cos \theta
\end{eqnarray}
and used notations  $b \equiv k \: \sin\theta, \;\; k \equiv
-\frac{v}{c}\frac{\sin\delta/2}{\delta/2}$. So k is a constant,
not a wave vector.  \\
 \indent The next step is to calculate the integral over
$\phi$. Because \cite{gr1965},
\begin{eqnarray}
\int_0^{2\pi}d\phi \frac{1}{(1+b \sin \phi)^4}=\frac{\pi(2 + 3
b^2)}{(1-b^2)^{7/2}},
\end{eqnarray}
\begin{eqnarray}
\int_0^{2\pi}d\phi \frac{\sin \phi}{(1+b \sin \phi)^4}=
\frac{-b\pi (4+b^2)}{(1-b^2)^{7/2}},
\end{eqnarray}
and
\begin{eqnarray}
\int_0^{2\pi}d\phi \frac{\sin^2 \phi}{(1+b \sin \phi)^4}=
\frac{\pi(1+4b^2)}{(1-b^2)^{7/2}},
\end{eqnarray}
the correlation function takes the form (\ref{eq-CF11Final}).
\section{The second way to receive the $\langle E_{(1)}(\mu_1|\tau_1)
E_{(1)}(\mu_2|\tau_2)\rangle$ for the Discrete Spectrum. }
\label{sec-ModifiedExpressionRandom}  In the section
(\ref{sec:CFDiscrete}) we have obtained the general expression for
the CF $\langle E_{(1)}(\mu_1|\tau_1)
E_{(1)}(\mu_2|\tau_2)\rangle$ for the discrete spectrum. \\
\indent Here we show that it can be received in a different way.
For the discrete spectrum the equations (\ref{eq:ff}) should be
 modified to:
\begin{eqnarray}
\vec{E}(\vec{r},t)= a\:\sum^{\infty}_{n=0} \sum^2_{\lambda=1} \int
do\,k^2_n\,\hat{\epsilon}(\hat{k},\lambda)\,h_0(\omega_n)\
\cos[\vec{k_n}\vec{r}-\omega_n
t -\Theta(\vec{k_n},\lambda)], \nonumber \\
\vec{H}(\vec{r},t)=a\:\sum^{\infty}_{n=0}\sum^2_{\lambda=1} \int
do \,k^2_n \,
[\hat{k},\hat{\epsilon}(\hat{k},\lambda)]\,h_0(\omega_n)\,\cos[\vec{k_n}\vec{r}-\omega_n
t-\Theta(\vec{k_n},\lambda)], \nonumber \\
 \vec{k}_n=k_n \hat{k},
\;\; k_n=k_0\,n, \;\; k_0= \frac{\Omega}{c}, \;\; \omega_n=c
\,k_n, \;\;
 do= d\theta \, d\phi \, \sin\theta, \nonumber \\
 \hat{k}=(\hat{k}_x,
\hat{k}_y, \hat{k}_z)=(\sin\theta \, \cos \phi, \, \sin\theta \,
\sin\phi, \,\cos\theta\,), \;\; a = k_0. \label{eq:mff}
\end{eqnarray}
 The unit vector $\hat{k}$ defines a
direction of the wave vector and does not depend on its value,
n.\\
The right side of the first equation in the relation
(\ref{eq:thetas}) should be modified. We do it in two steps. First
we rewrite them in a spherical momentum space  \cite{davydov1968},
p.656 as :
\begin{eqnarray}
\label{eq:cosaverage} \langle \cos\theta(\vec{k}_1 \lambda_1
)\cos\theta(\vec{k}_2 \lambda_2 ) \rangle=\langle
\sin\theta(\vec{k}_1 \lambda_1 )\sin\theta(\vec{k}_2 \lambda_2 )
\rangle=\frac{1}{2}\delta_{\lambda_1
\lambda_2}\delta^3(\vec{k}_1-\vec{k}_2)=
\frac{1}{2}\delta_{\lambda_1 \, \lambda_2} \, \frac{2}{k_1^2}\,
\delta(k_1-k_2)\delta(\hat{k}_1-\hat{k}_2).
\end{eqnarray}
And then, in the case of the discrete spectrum, it  takes the
form:
\begin{eqnarray}
\label{eq:cosaverage} \langle \cos\theta(\vec{k}_{n_1} \lambda_1
)\cos\theta(\vec{k}_{n_2} \lambda_2 ) \rangle= \langle
\sin\theta(\vec{k}_{n_1} \lambda_1 )\sin\theta(\vec{k}_{n_2}
\lambda_2 )\rangle=  \frac{1}{2}\delta_{\lambda_1 \, \lambda_2} \,
\frac{2}{k_0(k_0 n_1)^2}\, \delta_{n_1 \,
n_2}\delta(\hat{k}_1-\hat{k}_2).
\end{eqnarray}
The equation
 $\;\; \sum^2_{\lambda=1}\epsilon_i(\vec{k}\lambda
)\epsilon_j(\vec{k}\lambda )= \delta_{ij}-\hat{k}_i\hat{k}_j \;\;$
does not depend on n. The correlation function finally takes the
form (\ref{eq:CF11d1},\ref{eq:CF11d2})
\section{Another expression for $S_d $}
\label{sec-anotherExpressionSd} The sum $S_d$ has been calculated
in (\ref{eq:I_one}). It can also be given in a different form:
\begin{eqnarray}
\label{eq:I_two}
 S_d= \frac{6}{F_d^4} +
6  \sum_{n=1}^{\infty} \frac{1}{(2\pi n)^4}\;\; [\;\; \frac{1}
{(1+ F_d /2\pi n)^4}
  + \frac{1}{(1- F_d /2\pi n)^4} \;\; ].
\end{eqnarray}

\end{document}